\documentclass[a4paper, 11pt]{article}

\usepackage{a4wide}

\usepackage[english]{babel}
\usepackage[utf8]{inputenc}
\usepackage[T1]{fontenc}

\usepackage[ttscale=.875]{libertine}
\usepackage[libertine]{newtxmath}
\usepackage[scaled=1]{inconsolata}

\usepackage{graphicx}

\usepackage{booktabs}

\usepackage[backend=biber]{biblatex}
\title{Anomaly Detection in Certificate Transparency Logs}
\author{
  Richard Ostertág\qquad Martin Stanek \\[2ex]
  Department of Computer Science \\
  Faculty of Mathematics, Physics and Informatics \\
  Comenius University \\
  \textsl{\{richard.ostertag, martin.stanek\}@fmph.uniba.sk}
}

\begin{document}
\maketitle

\begin{abstract}
We propose an anomaly detection technique for X.509 certificates utilizing Isolation Forest. This method can be beneficial when compliance testing with X.509 linters proves unsatisfactory, and we seek to identify anomalies beyond standards compliance. The technique is validated on a sample of certificates from Certificate Transparency logs.
\end{abstract}

\section{Introduction}

Digital certificates, or public key certificates, issued by trusted certification authorities play an essential role in facilitating trust in security protocols. They bind the identity of a subject to a specific public key. Certificates that are issued mistakenly or with malicious intent pose a significant security threat, with impacts related to identity spoofing.

Certificate Transparency (CT) is a standard designed to mitigate this threat. The main idea behind CT is to collect and store all issued certificates in publicly available CT logs with verifiable authenticity. These logs allow anyone, such as domain owners, to monitor issued certificates and detect misissued certificates.
The details of CT operation, including participants, data structures, protocol, etc., are specified in RFC 9162~\cite{rfc9162}.

Certificate Transparency is gradually gaining popularity, and browsers like Chrome (Chromium) and Safari are now requiring Transport Layer Security (TLS) certificates to contain proof of CT log inclusion. This requirement is achieved by adding signed certificate timestamps (SCTs) into the certificate. The SCT serves as a signed promise that the CT log operator will append the certificate to the CT log.

The most prominent public CT logs are operated by Google, Cloudflare, and certification authorities themselves, such as DigiCert, Let's Encrypt, and Sectigo. Since all relevant certification authorities support CT, as of May 2024, over 460,000 certificates are published in CT logs every hour \cite{Cloudflare}.

The HTTP-based API that allows direct access to a CT log is specified in RFC 9162~\cite{rfc9162} (version 2.0) or RFC 6962~\cite{rfc6962} (version 1). However, the API is focused on monitoring CT log entries and there is no method to search for entries based on domain names or other attributes. To satisfy the demand for advanced queries and monitoring, there are various free and commercial services available. Notable free search services are \texttt{crt.sh}\footnote{\url{https://crt.sh}, a direct SQL access to the database is also available} operated by Sectigo and Entrust Certificate Search\footnote{\url{https://ui.ctsearch.entrust.com/ui/ctsearchui}}. Commercial offerings allow outsourcing monitoring tasks for domain owners and provide automated checks and notifications when events that require owner attention are observed.

In the world of ubiquitous Transport Layer Security (TLS) communication, CT logs have become a rich source of information regarding domain names. Passive reconnaissance regularly employs searches through CT logs to enumerate subdomains during penetration testing. Example tools that use this technique, among other methods, are OWASP Amass\footnote{\url{https://owasp.org/www-project-amass/}}, subfinder\footnote{\url{https://github.com/projectdiscovery/subfinder}}, and reconFTW\footnote{\url{https://github.com/six2dez/reconftw}}.

\paragraph{Anomaly detection.} Anomalous certificates may indicate various issues, such as misissued certificates, unintended defects, or operational problems of domain owners. They can raise suspicions and warrant an investigation. Certificates in CT logs can even be abused for unidirectional covert communication~\cite{Jurcak23}. There might be other abuses of CT logs and unknown problems as well. It is much more efficient to detect misissued certificates using exact tests when we know what we are looking for. However, the detection of anomalous certificates can help identify potential, yet unknown, issues that may require further investigation.

Another application of anomaly detection is when anomalies initially identified by a model are no longer rare. This might indicate changes in the use of certificates, reflected in their structure or content characteristics. Moreover, the model can be trained on certificates issued for specific domains, and anomalies detected in newly issued certificates can indicate an internal problem that needs to be addressed. 

In our paper, we use the term ``anomaly'' to refer to certificates that are significantly different from those usually observed. We do not test certificates for compliance with X.509 standards like linters do\footnote{It is important to note that X.509 linters check certificates against a specific set of rules, ensuring they conform to established standards. Some well-known tools are ZLint and pkilint.}. However, in future work, it might be interesting to include linter results as additional attributes for anomaly detection, providing a more comprehensive analysis of certificate structures and content. 

\paragraph{Our contribution.} We evaluate selected statistical information about certificates in CT logs, focusing on attributes defined by domain owners, such as Subject Alternative Name (SAN) in Section \ref{sec-stats}. We propose a method for anomaly detection in certificates using Isolation Forest \cite{IF08,IF12}, an unsupervised machine learning technique, in Section \ref{sec-anom}. We select suitable certificate attributes and train the model on a sampled set of certificates obtained from CT logs. The results of our Isolation Forest model are presented in Section \ref{sec-res}.

\section{Statistics and attributes selection}
\label{sec-stats}

We created a random sample of 120,000 records from one of the largest public Certificate Transparency (CT) logs, Xenon 2024, which is operated by Google. In CT logs, there are two types of records: precertificates and certificates. Since all the features we want to extract are already available in precertificates, we do not discriminate between these types in our analysis.

According to the statistics presented by Cloudflare on their Merkle Town webpage \cite{Cloudflare}, the issuance rate of new certificates across all monitored CT logs is more than 460 thousands per hour (as of May 2024). Therefore, the probability of sampling both the corresponding precertificate and certificate is negligible. The sample size and variety of records in our experiment are sufficient to effectively investigate anomalous certificates within CT logs.

Let us discuss what attributes we considered and selected for feature extraction. We group them in several categories -- subject, subject's public key, issuer, signature, validity, and X.509 extensions.

\paragraph{Subject.} A distinguished name (DN) consists of a set of attributes that identify a subject. In the case of domain validated certificates, it usually contains just the common name (CN). For organization validated certificates, however, it may contain a set of attributes such as:

\begin{verbatim}
   CN=multimedia-academy.tudelft.nl,
   O=Technische Universiteit Delft,
   ST=Zuid-Holland,
   C=NL
\end{verbatim}

The presence and number of attributes in a DN can vary. For instance, we found that approximately $2.28$\% of our sample certificates did not contain a CN attribute. To extract quantitative features from the subject section of a certificate, we considered the following characteristics (see also a boxplot visualization in the Figure \ref{fig0-subject}):

\begin{figure}
\centering
\includegraphics[width=0.2\textwidth]{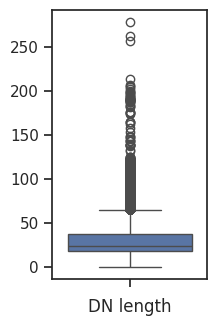}
\includegraphics[width=0.2\textwidth]{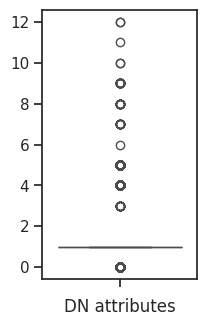}
\includegraphics[width=0.2\textwidth]{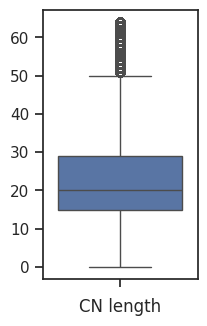}
\includegraphics[width=0.2\textwidth]{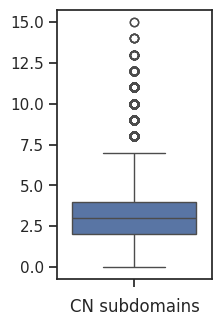}
\caption{Selected characteristics of subjects in the dataset}\label{fig0-subject}
\end{figure}

\begin{itemize}
\item The length of a DN -- this refers to the number of characters in the DN string representing a subject. In our sample, DN lengths range from $0$ to $278$ characters with an average length of $33.0$.
\item The number of attributes in a DN -- this represents the inner structure of the DN and indicates how many relative Distinguished Names (RDNs) are present. The maximum value in our sample is $12$ attributes, while the mean is $1.4$ attributes, and only $14.0$\% of records have an attribute count that is not equal to $1$.
\item The length of a CN -- this attribute focuses on the most important and most frequently present part of a DN. The maximal allowed length of $64$ characters \cite{rfc5280} is observed in $1.8$\% of records.
\item Number of subdomains in a CN -- this represents the inner structure of CN. In our sample, the number of subdomains ranges from $0$ to $15$.
\item Wildcard CN -- a boolean value indicating whether the CN contains a `\texttt{*}' character. Wildcard CNs are observed in $12.0$\% of records.
\end{itemize}

Certainly, there might exist qualitative anomalies in certificates based on small differences or variances that are not captured by quantitative characteristics alone. For example, some uncommon semantics may be used for DN attributes. These anomalies will not be detected with methods trained only on quantitative features. However, we do not attempt to analyze these anomalies in this paper as it would require interpreting different parts and attributes of the certificate beyond the scope of our experiment. This approach, focusing on quantitative characteristics, is also used for feature extraction in the rest of this section.

\paragraph{Subject's public key.} A public key is another attribute that is fully controlled by the subject. The certificate authority can restrict the types and supported lengths of public keys for issued certificates, but the value is ultimately generated by the subject. We extract two features from the public key: its type and length.

\begin{itemize}
\item Public Key Type: There are only two types of subject public keys – RSA and Elliptic Curve Digital Signature Algorithm (ECDSA). Our sample shows a dominant position of RSA keys ($73.5$\%). We do not extract the type of elliptic curve used in ECDSA keys. A numeric encoding of public key type is performed as follows: ECDSA $\mapsto 0$, RSA $\mapsto 1$.
\item Public Key Length: Bit length of the public key, depending on the modulus length for RSA or chosen curve for ECDSA. The observed variability of this attribute is presented in Table \ref{tab-pklengths}.  
\end{itemize}

\begin{table}
\begin{center}
\begin{tabular}{@{}lrrrrrr@{}}
        & $256$ & $384$ & $2048$  & $3072$ & $4096$ & $8192$ \\
  \midrule
  RSA   & & & $64.8$\% & $0.4$\% & $8.4$\% & $0.0\text{\%}^{\text *}$ \\
  ECDSA & $24.4$\% & $2.1$\% & & & &  \\[2ex]
  \multicolumn{7}{l}{* {\small exactly one $8192$-bit RSA key in the sample}}
\end{tabular}
\end{center}
\caption{Distribution of subject's public key lengths in the sample}\label{tab-pklengths}
\end{table}

\paragraph{Issuer.} Let's Encrypt is the most prevalent certification authority, accounting for over $52$\% of certificates in our sample. The total number of distinct certification authorities, identified by unique DN, is $176$. For anomaly detection, we will use the rarity of CA as a feature:

\begin{itemize}
\item CA rarity: A float number computed as a fraction of certificates in the sample with the same issuer (DN).
\end{itemize}

It is assumed that more common certification authorities have better practices and stricter certification policies in place, so their certificates are less likely to be anomalous. Therefore, we will not analyze other aspects of the issuer further.
 
\paragraph{Signature.} We do not extract any features from a signature algorithm used by certification authorities to sign (pre)certificates. This is entirely at their discretion, and we assume that CA rarity, see above, covers unusual certification authorities sufficiently in our experiment. However, if someone wants to consider signatures in anomaly detection, both types (algorithms) as well as key lengths should be considered. Nice online statistics covering signature algorithms are presented in \cite{Cloudflare}, with RSA-SHA256 being used in $90$\% of (pre)certificates.

Another set of attributes that might be considered in the future are embedded SCT (Signed Certificate Timestamps) in certificates. A certification authority can decide in which CT logs it wants to include a certificate. This decision is usually uniform across different certificates, taking into account their expiry date. An unusual combination or SCT count can indicate an anomaly.

\paragraph{Validity.} Despite the validity period depends on the CA's certification policy, our sample demonstrates significant variability within this attribute, ranging from one day to approximately $50$ months. This feature is extracted for use in anomaly detection.

\begin{itemize}
\item Validity period: the number of days a certificate is valid, calculated as the difference between ``not before'' and ``not after'' dates. Approximately $70$\% of certificates in our sample are issued for a validity period of three months, predominantly due to Let's Encrypt's certification policy. Nearly $19.3$\% of the certificates have a validity period of approximately one year.
\end{itemize}

\paragraph{X509 extensions.}

There are various extensions that can be part of a certificate. Our experiment with anomaly detection is focused mostly on attributes chosen by the subject. Therefore, special attention is given to the features of Subject Alternative Name (SAN) extension. According to RFC 5280 \cite{rfc5280}, the SAN entry can contain DNS names, IP addresses, internet electronic mail addresses, Uniform Resource Identifiers (URIs), and other options exist as well. The sample shows an overwhelming probability of DNS names, where almost all certificates have at least one DNS name in the SAN extension. Other entry types appear in negligible fractions of records: IP addresses are present in less than $0.03$\% of records, and other types are absent altogether. We extract the following features for anomaly detection:

\begin{figure}[h]
\includegraphics[width=0.5\textwidth]{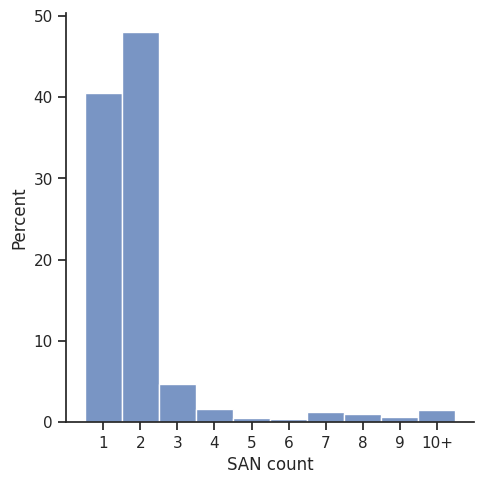}
\includegraphics[width=0.5\textwidth]{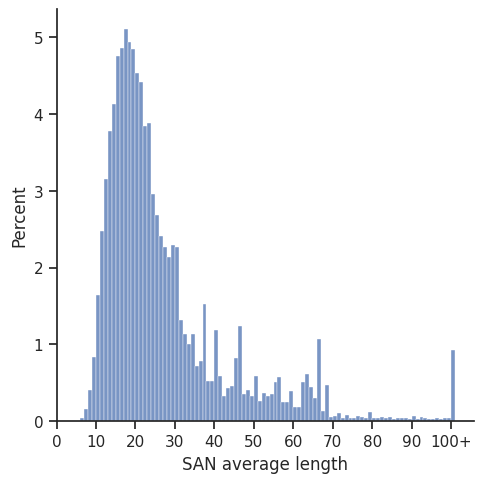}
\caption{SAN count and average length in the dataset}\label{fig-san}
\end{figure}

\begin{figure}[h]
\centering\includegraphics[width=0.5\textwidth]{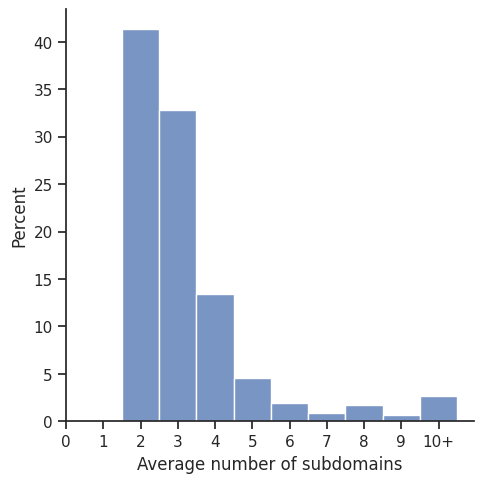}
\caption{Average number of subdomains in the dataset}\label{fig-san2}
\end{figure}

\begin{itemize}
\item The count of SAN entries: Our sample shows an average number of SAN entries as $2.1$, with a minimum of $1$ and a maximum of $238$. The number of certificates with $10$ or more SANs is below $1.5$\%.
\item Average length of SAN entries: The average length of SAN entries in a certificate ranges from $5$ to $239$ with an average value of $27.3$. Given the observation of SAN count, the value primarily depends on certificates with a small number of SAN entries. The distribution of average SAN length values along with the distribution of SAN count values is presented in Figure \ref{fig-san}.
\item The number of wildcard domain names: Approximately $65$\% of the certificates do not contain any wildcard names in their CN and SAN attributes, while $31.1$\% of certificates have just one wildcard name. Other counts are significantly less represented (less than $3.9$\%).
\item Average number of subdomains: The average number of subdomains for CN and SAN attributes is calculated by counting all substrings separated by periods (``.'') in a domain name. For example, ``www.uniba.sk'' has three subdomains: ``www'', ``uniba'', and ``sk''. As expected, the average number of subdomains is generally within the range of 2 to 4, as shown in Figure~\ref{fig-san2}.
\item Validation type: We assign each certificate a numerical representation of its validation type, with 0 representing missing or unavailable information, 1 for Domain Validation (DV), 2 for Organizational Validation (OV), and 3 for Extended Validation (EV). This representation orders validation types from the least strict policy to the most strict validation policy. For comprehensive global statistics, see Merkle Town's webpage \cite{Cloudflare}. In our sample, we found that $88.3$\% of certificates were DV, and $11.7$\% were OV, while other types occurred negligibly.
\end{itemize}

We decided not to analyze other extensions separately despite their potential interest, such as Key Usage, CRL, OCSP, and various constraints. Although problems or anomalies can be hidden in any of them, we selected a subset of attributes more related to the subject, because these attributes can help detect incorrect configurations when requesting certificates or possible covert communication. For other anomaly detection applications, it might be important to include specific X509 extensions in the set of selected features. Our experiment focuses on the following summary characteristics:

\begin{itemize}
\item Extensions count: The number of X509 extensions in a certificate. The dataset shows this parameter ranging from $5$ to $13$ with $97.3$\% of records having $9$ or $10$ extensions.
\item Extensions size: The length of X509 extensions in a certificate excluding SAN, since the related SAN characteristics -- number and average length -- are represented separately. The average size in our sample is $2306$ bytes, while minimum and maximum sizes are $815$ and $3506$ bytes, respectively.
\end{itemize}

\section{Anomaly detection}
\label{sec-anom}

Isolation Forest is an unsupervised anomaly detection technique proposed by Liu, Ting, and Zhou \cite{IF08,IF12}. It builds a collection of binary trees, similar to binary search trees, by randomly selecting branching features and thresholds. The anomaly score for a data point is based on the average depth at which it is isolated across multiple trees. The main idea behind Isolation Forest is that, on average, anomalies are isolated in lower depths than non-anomalous data.

The Isolation Forest algorithm was selected for our experiment due to its ability to detect anomalies without relying on complex distance metrics or density estimation. Furthermore, Isolation Forest performs well in high-dimensional problems containing a large number of irrelevant attributes. Additionally, it can effectively train the model even when the anomalies are not present in the training sample. The technique also has low time and memory complexity.

We utilize an implementation of the Isolation Forest provided in PyOD library \cite{zhao2019pyod} for anomaly detection in multivariate data. We set the following parameters for this technique:

\begin{itemize}
\item Number of estimators (trees): $200$
\item Number of samples drawn from the data to train each estimator: $256$
\item Number of features drawn from the data to train each estimator: $16$ (all available features)
\item Sampling from the data is performed without replacement.
\end{itemize}

The contamination of the data, i.e. the proportion of anomalies in the dataset, is irrelevant for the discussion in Section \ref{sec-res}. The reason being that the contamination is only used to set an anomalous score threshold. Instead, we examine which data, specifically precertificates and certificates, have the highest anomalous scores. From these observations, conclusions can be drawn without requiring knowledge of the exact contamination value for our dataset.

\section{Results}
\label{sec-res}

We document the types of precertificates and certificates that are detected as the most anomalous in our exploratory experiment. A general observation is that some cloud services and their internal components are the most frequent outliers in our dataset.

\paragraph{Azure infrastructure.}

The most anomalous certificates in our experiment are those issued by Microsoft for the components of Azure infrastructure. The issuing CAs are:

\begin{itemize}
\item Microsoft Azure TLS Issuing CA \textsl{XX} -- several authorities, where \textsl{XX} denotes number 01, 02, etc.;
\item Microsoft Azure RSA TLS Issuing CA \textsl{XX} -- again several authorities issuing certificates;
\item Microsoft RSA TLS CA \textsl{XX} -- significantly smaller number of certificates in comparison to the above two sets of authorities. 
\end{itemize}

\begin{table}[h]
\begin{center}
\begin{tabular}{@{}lrrrr@{}}
     & DN                & CN     & SAN          & extensions \\
  CA & attributes/length & length & count/length & count/size \\
  \midrule
  Microsoft Azure TLS Issuing CA & 5.0/89.6 & 41.6 & 3.5/45.0 & 12.0/3218 \\
  Microsoft Azure RSA TLS Issuing CA & 5.0/97.3 & 49.3 & 3.4/54.5 & 12.0/3232 \\
  Microsoft RSA TLS CA & 1.0/38.5 & 35.5 & 21.2/38.2 & 10.7/3031
\end{tabular}
\end{center}
\caption{Averages for selected characteristics of issued certificates}\label{tab-azure}
\end{table}

Table \ref{tab-azure} summarizes basic characteristics for each CA. It shows above-average values, particularly for the first two CAs. Besides higher than usual number of SAN domain names, longer domain names and extensions, the other factors contribute to anomaly of detected certificates as well. Top anomalous certificates show various deviations, such as slightly odd validity period, the number of wildcard domain names, and other attributes, combined with relative rarity of issuing CA. In this regard, the anomaly detection works as intended. For example, the most anomalous certificate in the dataset according our trained model shows the following characteristics: 
\begin{itemize}
\item Common Name: \texttt{CN=*.table.preprod.core.windows.net}
\item Issuer: Microsoft Azure TLS Issuing CA 06
\item Validity period: $282$
\item SAN count: $52$, the number of wildcard domain names: $52$
\item The average number of subdomains: $7$
\item The number of extensions: $12$, overall extension size: $3206$
\end{itemize}

\paragraph{Other CAs and ZeroSSL.} 

After filtering out certificates issued by the CAs metioned above, we examined the top 100 anomalous items in greater detail. Among these, we observed:

\begin{itemize}
\item Two certificates issued by DigiCert: one for a Chinese cloud service provider and one for a legitimate IT company.
\item Two certificates issued by Amazon for its AWS components.
\item All remaining 96 certificates were issued by ZeroSSL CA, specifically by ZeroSSL ECC Domain Secure Site CA. These have an unusual structure: empty subject (DN), and questionable SAN attributes containing a large number of repetitive subdomains. Two examples are:
  \begin{itemize}
  \item \texttt{www.www.www.www.www.www.pay.avito.sber.avito.avito.www.www.www.}\\
  \texttt{www.www.www.www.www.yandex.avito.yandex.pay.portalswebmail.}\\
  \texttt{blumebwww.od3.10cekub2b.k.webmail.ultagkhanub2b.k.webmemo.m.}\\
  \texttt{phpmyadmin.wokemtutankhanub2b.k.webmail.ultagame.com}
  \item \texttt{www.www.www.www.www.www.www.www.www.www.www.www.www.www.www.www.}\\
  \texttt{www.www.www.www.www.www.www.www.www.www.www.www.www.www.www.www.}\\
  \texttt{www.www.www.www.www.www.www.www.www.www.www.www.www.www.www.www.}\\
  \texttt{www.www.calendario.panel-fiveheberg.fr}
  \end{itemize}

These might indicate an operational problem with an automation script that issues and renews certificates and adds \texttt{www} prefix to a domain name. Additionally, in case of the domain \texttt{panel-fiveheberg.fr}, combined with a wildcard DNS record that positively responds to any DNS query. We checked both domains in VirusTotal, and no security vendor flagged them as malicious (as of April 2024).

  Not all precertificates and certificates issued by ZeroSSL ECC Domain Secure Site CA in our dataset have the mentioned structure. However, the fraction with an empty subject is rather large: $41.3$\%. Table \ref{tab-zero} summarizes some characteristics of records issued by this CA. It's an interesting observation that free certificates issued by Let's Encrypt CA do not exhibit such anomalies\footnote{There are only 7 (pre)certificates with empty subject out of 62,424 issued by any Let's Encrypt CA in our dataset.}.
\end{itemize}

\begin{table}[h]
\begin{center}
\begin{tabular}{@{}lrrrr@{}}
      & DN                & CN     & SAN          & extensions \\
  set & attributes/length & length & count/length & count/size \\
  \midrule
  all (pre)certificates & 0.6/18.9 & 17.1 & 1.0/61.2 & 9.0/2307 \\
  empty subject & 0.0/0.0 & 0.0 & 1.0/106.7 & 9.0/2307
\end{tabular}
\end{center}
\caption{Averages for (pre)certificates issued by ZeroSSL ECC Domain Secure Site CA}\label{tab-zero}
\end{table}

The problem with unusual length of the CN attribute is not unique to ZeroSSL CA. Similar certificates are issued by Let's Encrypt. Again, possible explanation might be an error in certificate management automation. Domain owners are probably not aware or simply do not care, since both CA offers free certificates. An examples of such CN (certificate issued by Let's Encrypt) is

\begin{center}
\texttt{gitlab.gitlab.gitlab.gitlab.gitlab.git.git.testing.yikj.work}
\end{center}

\paragraph{Other observations.}

Ignoring ZeroSSL-issued certificates and various additional infrastructure certificates by Apple, Cisco, Google, and other well-known companies, we have found several more entries that look interesting. For example, a certificate issued by Let's Encrypt CA with the following set of SANs:
\begin{center}
\begin{tabular}{l}
\texttt{*.ajptzd.com, *.amklvv.com, *.aqcssg.com, *.ccjytp.com, *.doeigp.com,}\\
\texttt{*.egfnjv.com, *.eydqoa.com, *.fvrnlf.com, *.guuzxk.com, *.hgmwfy.com,}\\
\texttt{*.iwhqyn.com, *.kldcuc.com, *.lfmdnj.com, *.lloond.com, *.naktki.com,}\\
\texttt{*.nmklqi.com, *.npwpbz.com, *.nxezmi.com, *.ojdger.com, *.psfqpu.com,}\\
\texttt{*.ptgreh.com, *.raclbc.com, *.rvaajo.com, *.spikfh.com, *.swwoyd.com,}\\ 
\texttt{*.tnuntp.com, *.xfcpkw.com, *.xnrsre.com, *.xuvvdq.com, *.yyiosx.com}
\end{tabular}
\end{center}

\noindent Most of these domains are unresolvable by public DNS (NXDOMAIN) as of May 2024. We did not investigate this certificate further, whether it is a case of legitimate use-case, misconfiguration, business malpractice, or other malicious intent.

\section{Conclusion}

We proposed an anomaly detection technique for certificates using Isolation Forest. This approach can be beneficial when compliance testing with X.509 linters is unsatisfactory, and we seek anomalies beyond compliance. We demonstrated the feasibility of this method; however, further exploration is necessary. Some potential directions are:

\begin{itemize}
\item Training the model on certificates for a specific domain or domains owned by a single entity, allowing anomalies to serve as early internal warnings of potential issues.
\item Identifying certificates from large cloud providers and excluding them from the model and evaluation. The CT logs contain a vast quantity of these precertificates and certificates, which can distort parameters of the model.
\item Analyzing the results of identified anomalies in greater detail, such as those described in the previous section, to find explanations for the anomalous certificates.
\end{itemize}

\subsection*{Acknowledgment}
This publication is the result of support under the Operational Program Integrated Infrastructure for the project: Advancing University Capacity and Competence in Research, Development a Innovation (ACCORD, ITMS2014+:313021X329), co-financed by the European Regional Development Fund.

\printbibliography

@misc{rfc6962,
    series =    {Request for Comments},
    number =    6962,
    howpublished =  {RFC 6962},
    publisher = {RFC Editor},
    doi =       {10.17487/RFC6962},
    url =       {https://www.rfc-editor.org/info/rfc6962},
    author =    {Ben Laurie and Adam Langley and Emilia Kasper},
    title =     {{Certificate Transparency}},
    pagetotal = 27,
    year =      2013,
    month =     jun,
}

@misc{rfc9162,
    series =    {Request for Comments},
    number =    9162,
    howpublished =  {RFC 9162},
    publisher = {RFC Editor},
    doi =       {10.17487/RFC9162},
    url =       {https://www.rfc-editor.org/info/rfc9162},
    author =    {Ben Laurie and Adam Langley and Emilia Kasper and Eran Messeri and Rob Stradling},
    title =     {{Certificate Transparency Version 2.0}},
    pagetotal = 53,
    year =      2021,
    month =     dec,
}

@mastersthesis{Jurcak23,
    author  = {Matej Jurčák},
    title   = {Using Certificates and CT Logs for communication},
    school  = {Comenius University},
    year    = 2023,
    type={Bachelor's Thesis},
    note    = {In Slovak},
}

@inproceedings{IF08,
    author={Liu, Fei Tony and Ting, Kai Ming and Zhou, Zhi-Hua},
    booktitle={2008 Eighth IEEE International Conference on Data Mining}, 
    title={Isolation Forest}, 
    year={2008},
    volume={},
    number={},
    pages={413-422},
    doi={10.1109/ICDM.2008.17}}

@article{IF12,
    author = {Liu, Fei Tony and Ting, Kai Ming and Zhou, Zhi-Hua},
    title = {Isolation-Based Anomaly Detection},
    year = {2012},
    issue_date = {March 2012},
    publisher = {Association for Computing Machinery},
    address = {New York, NY, USA},
    volume = {6},
    number = {1},
    issn = {1556-4681},
    doi = {10.1145/2133360.2133363},
    journal = {ACM Trans. Knowl. Discov. Data},
    month = {mar},
    articleno = {3},
    numpages = {39}}

@online{Cloudflare,
    title        = {Merkle Town},
    author    = {Cloudflare},
    year       = 2023,
    url          = {https://ct.cloudflare.com/},
}

@misc{rfc5280,
    series =    {Request for Comments},
    number =    5280,
    howpublished =  {RFC 5280},
    publisher = {RFC Editor},
    doi =       {10.17487/RFC5280},
    url =       {https://www.rfc-editor.org/info/rfc5280},
    author =    {Sharon Boeyen and Stefan Santesson and Tim Polk and Russ Housley and Stephen Farrell and David Cooper},
    title =     {{Internet X.509 Public Key Infrastructure Certificate and Certificate Revocation List (CRL) Profile}},
    pagetotal = 151,
    year =      2008,
    month =     may,
 }

@article{zhao2019pyod,
    author  = {Zhao, Yue and Nasrullah, Zain and Li, Zheng},
    title   = {PyOD: A Python Toolbox for Scalable Outlier Detection},
    journal = {Journal of Machine Learning Research},
    year    = {2019},
    volume  = {20},
    number  = {96},
    pages   = {1-7},
    url     = {http://jmlr.org/papers/v20/19-011.html}
}

\end{document}